\documentclass[aip,amsmath,amssymb,reprint,floatfix]{revtex4-1}
\usepackage{graphicx}
\usepackage{fixltx2e}
\usepackage{dcolumn}
\usepackage{bm}
\usepackage[utf8]{inputenc}
\usepackage[T1]{fontenc}
\usepackage{xcolor}
\usepackage{soul}

\usepackage[normalem]{ulem}
\usepackage{soul,xcolor}
\usepackage{mathptmx}
\usepackage{amsmath}
\usepackage[colorlinks=true,citecolor=blue]{hyperref}%
\usepackage{subcaption}
\usepackage{gensymb}
\usepackage{upgreek}
\usepackage{lineno}
\usepackage{lipsum}
\draft 

\begin{document}
%


\title[Cold Atmospheric Microplasma Jet–Water Interactions: Physicochemical Analysis...]{Cold Atmospheric Microplasma Jet–Water Interactions: Physicochemical Analysis and Growth Effects in Flowering Plants} 

\author{Syon Bhattacharjee}
\email{bsyon2108@gmail.com}
 \affiliation{Delhi Public School Kalyanpur, Kanpur-208017, India}  
\author{Deepika Behmani}
\email{deepikabehmani28@gmail.com}
\affiliation{Department of Physics, Indian Institute of Technology Kanpur, Kanpur-208016, India}%
\author{Sudeep Bhattacharjee}
\email{sudeepb@iitk.ac.in}
\affiliation{Department of Physics, Indian Institute of Technology Kanpur, Kanpur-208016, India}





\date{\today}

\begin{abstract}
Cold atmospheric pressure plasma jets (APPJs) are non-equilibrium plasmas, that are capable of producing reactive oxygen and nitrogen species (RONS) at near-room temperature. Their interaction with water leads to the formation of plasma-activated water (PAW), whose chemical activity depends on discharge conditions. In this work, a helium-air (14:1) micro-plasma jet operated in a ring-to-ring electrode configuration is used to generate PAW and study its influence on the growth of Chrysanthemum saplings. Optical emission spectroscopy (OES) confirms the presence of $N_2$ bands and He lines, with the He-air mixture providing more chemically active discharge (in terms of favoring the generation of nitrates in PAW) as compared to pure helium. The physicochemical characteristics of PAW such as pH, electrical conductivity (EC), oxidation-reduction potential (ORP), and total dissolved solids (TDS) are analyzed as a function of plasma treatment time and water volume. The optimum condition for PAW generation is found to be 12 ml of de-ionized (DI) water treated for 40 minutes, which yields the highest ORP and nitrate concentration with a reduced pH. Comparative growth experiments over two weeks show that PAW-treated Chrysanthemum saplings exhibit significantly greater height (10.2 cm) and soil fertility (2580 µS/cm) than those watered with same amount of DI water or tap water. The results highlight the potential of PAW for sustainable enhancement of growth of flowering plants.
\end{abstract}


\maketitle 


\section{Introduction}
Cold atmospheric pressure plasma jets (APPJs) have gained significant attention over the past few decades due to their ability to generate highly reactive, non-equilibrium plasmas at near-room temperature and atmospheric pressure \cite{bardos2010cold}. Unlike conventional thermal discharges, APPJs operate in a regime having a significantly different electron and ion temperature, which facilitates the occurrence of a wide range of chemical reactions without substantial gas heating. These unique features make APPJs suitable for diverse applications in fields such as material processing \cite{fanelli2017atmospheric}, biomedical research \cite{keidar2013cold,zimu2020applications,lin2021map,laroussi2009low}, cancer treatment \cite{zimu2020applications,keidar2013cold}, surface modification \cite{bornholdt2010characterization,kim2003surface,shaw2016mechanisms,mello2012surface,barman2023improving}, environmental \cite{bogaerts2022foundations}, food safety \cite{kim2015effect}, and agriculture \cite{rathore2024innovative, adhikari2019cold}.

Since APPJs operate in ambient air, they interact strongly with the environment, leading to the generation of a rich mixture of reactive oxygen and nitrogen species (RONS) such as $OH$, $O$, $NO$, $NO_2$, $H_2O_2$, and $O_3$ \cite{lin2021map}. These reactive species play a decisive role in downstream chemical processes, influencing both gas-phase and liquid-phase reactions \cite{wang2025plasma, rathore2024innovative,machala2018chemical,ogawa2018modulating,oh2019tailoring}. In recent years, there has been growing interest in exploiting these reactive species for plasma-assisted agriculture, an emerging field that integrates plasma science with plant physiology to promote sustainable growth and yield enhancement \cite{konchekov2023advancements,heping2022applications,guo2021plasma}.

One of the most promising tools in this domain is the use of plasma-activated water (PAW): the water treated with cold plasma discharges, which becomes enriched with reactive oxygen and nitrogen species after treatment \cite{rathore2024innovative, rathore2024effects, wang2025plasma}. The long-lived chemical species generated in PAW, such as nitrate ($NO_3^-$), nitrite ($NO_2^-$), and hydrogen peroxide ($H_2O_2$), are known to enhance seed germination, nutrient uptake, and root development \cite{adhikari2019cold, rathore2024innovative,barjasteh2023recent,sivachandiran2017enhanced, maniruzzaman2017nitrate}. Moreover, PAW offers an environmentally friendly alternative to chemical fertilizers \cite{rathore2024effects,pakprom2024optimizing,graves2019plasma} and can be easily stored and applied to plants, making it a scalable nutrient solution for agricultural practices.

Despite several studies exploring PAW for crop and vegetable growth, research on flowering plant systems remains limited \cite{guo2021plasma,guragain2023improvements,stoleru2020plant,carrillo2023finding}. The effect of gas composition and discharge parameters on PAW generation and its subsequent influence on plant development is not yet fully understood. Most reported works employ air, nitrogen, or oxygen discharges, which often suffer from instability and limited control over reactive species generation at atmospheric pressure. In contrast, helium-based discharges, particularly helium-air mixtures, provide enhanced discharge stability, improved energy transfer, and efficient production of RONS under ambient conditions when the helium-air mixture is properly optimised. \cite{behmani2021fluctuations,behmani2024frequency}. Flowering plants provide nectar, pollen, fruits and seeds and thus act as the foundation of the food web for the entire ecosystem. Moreover, the horticultural applications of flowering plants provide limitless opportunities. 

In this article, a helium-air cold atmospheric pressure micro-plasma jet operated in a ring-to-ring electrode configuration is employed to generate plasma-activated water and investigate its physicochemical properties and its effects on flowering plant growth. The study focuses on the growth enhancement of Chrysanthemum saplings, a representative flowering plant species that has not been explored in PAW-related research. The objectives of this work are to: (i) characterize the optical emission features of the plasma discharge to identify the active species generated in the He-air environment, (ii) analyze the variations in the physicochemical parameters and ionic composition of PAW with plasma treatment time, and (iii) evaluate the effect of optimized PAW on plant height and soil fertility. This systematic approach aims to establish a correlation between discharge characteristics, chemical activation of water, and response to plant growth, thereby contributing to the understanding and application of plasma-based agriculture.

\begin{figure*}
\centering
\includegraphics[scale = 0.6]{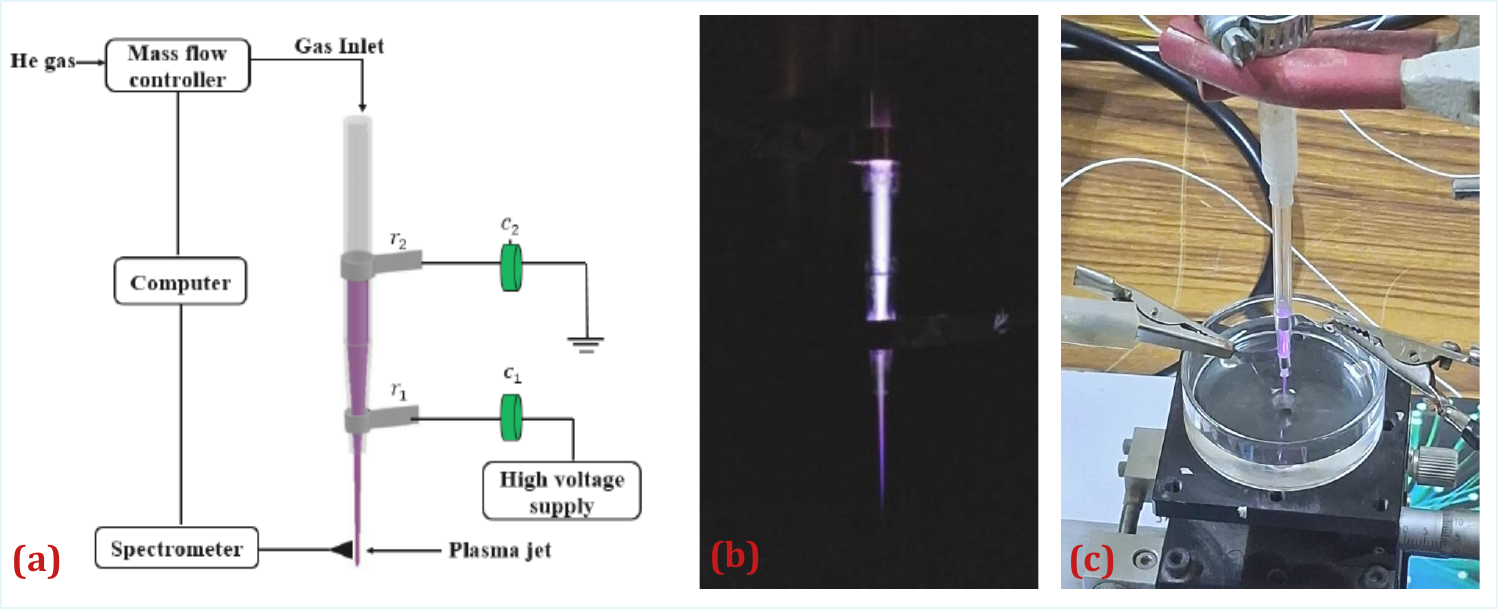}
\caption{(a) Atmospheric pressure plasma jet setup (ring-to-ring electrode configuration) \cite{barman2020characteristics}, (b) digital picture of plasma jet, and (c) digital picture of plasma activation of water}
\label{fig:1}
\end{figure*}
\normalsize

\section{Experimental Setup and Measuring Devices}

The experimental setup is shown schematically in Fig. \ref{fig:1}(a) \cite{barman2020characteristics, behmani2021fluctuations, behmani2023plasma, behmani2024frequency}, and a photograph of the plasma jet is shown in Fig. \ref{fig:1}(b). The discharge is produced inside a dielectric quartz capillary tube. The tube has an inner diameter (ID) of 2.5 mm and outer diameter (OD) of 4.14 mm at the inlet, tapering to an ID of 0.8 mm and OD of 2.06 mm at the outlet. The two ring electrodes that are placed 10 mm apart are made by wrapping 5 mm wide aluminum tapes around the capillary tube. The lower high-voltage (HV) electrode is positioned about 3 mm upstream from the capillary orifice to establish a strong electric field near the exit region.

A sinusoidal AC waveform of 10 kHz, having a peak to peak (pp) amplitude of 14 kV, and a discharge current of $\sim$25 mA is applied to ignite the working gas using a high voltage power supply (IONICS, model HV015K500WSIB4L). The working gas is a mixture of helium and air, which was optimized in a 14:1 ratio as it yielded the maximum concentration of nitrates within the water activation time. In accord with this ratio, Helium (3000 sccm) is the primary carrier gas and air (220 sccm) is an admixture. The flow rates of both the gases are precisely regulated using two different mass flow controllers (ALICAT Scientific, model MC-10SLPM-D-DB9M/5M). The addition of air enhances the chemical reactivity by introducing nitrogen and oxygen species essential for the formation of RONS. In contrast, pure helium discharge limits the production of nitrogen reactive species, inhibiting the formation of nitrates in liquid medium.

The plasma jet is directed vertically downward toward the surface of de-ionized (DI) water, which is placed in a glass Petri dish, for plasma activation, as shown in Fig. \ref{fig:1}(c). The distance between the capillary orifice (from where the plasma jet emerges) and the water surface is maintained at about 7 mm, ensuring efficient transfer of reactive species from the plasma to the liquid without causing significant heating or splashing.

Optical emission spectroscopy (OES) is used to qualitatively analyze discharge characteristics and identify active species. For this purpose, an Andor Technologies Shamrock 750 spectrometer with a resolution of 0.02 nm and an open electrode detector (DU420A-OE) are employed \cite{barman2020characteristics, barman2021effect, behmani2023plasma}. An optical fiber (SR-OPT 8024) having one channel with 19 UV/VIS fibers of 200 $\mu$m core diameter has been employed to collect the optical signal. Prominent emission lines of nitrogen ($N_2$) bands and helium ($He$) were observed in the discharge spectra, confirming the formation of a chemically-active environment suitable for plasma-liquid interaction.

After each plasma treatment, the activated water sample was subjected to physicochemical characterization: measurements of pH, electrical conductivity (EC), oxidation-reduction potential (ORP), and total dissolved solids (TDS) were carried out using a calibrated amiciSense Water Quality Tester. Ion chromatography (IC) confirmed the generation of nitrate ($NO_3^-$) species in the treated water samples. The 6 Chrysanthemum saplings chosen for our experiments were all of the same age and height. They were then kept in 3 miniature pots (2 in each) in a Seed Starter tray (Qoolife) with controllable plant light that mimicked natural sunlight and with
humidity domes; the pots were filled with a standard potting
soil mix of Red soil purchased from Trustbasket. Each of the 3 pots was labeled for treatment with tap water (TW), de-ionised water (DI) and plasma activated DI water (PAW). The saplings were daily watered with 12 ml of each type of water over a period of 2 weeks and the sapling growth was weekly recorded with a ruler scale. The average height of the two saplings in each pot was taken as the resultant height for that category of treated saplings. At the end of the two-week period, the fertility of TrustBasket potting soil was evaluated in each of the 3 pots using an amiciSense 6-in-1 Soil Tester. This allowed for a comprehensive assessment of how plasma treatment influences water chemistry and soil fertility.

\begin{figure}
\centering
\includegraphics[scale = 0.5]{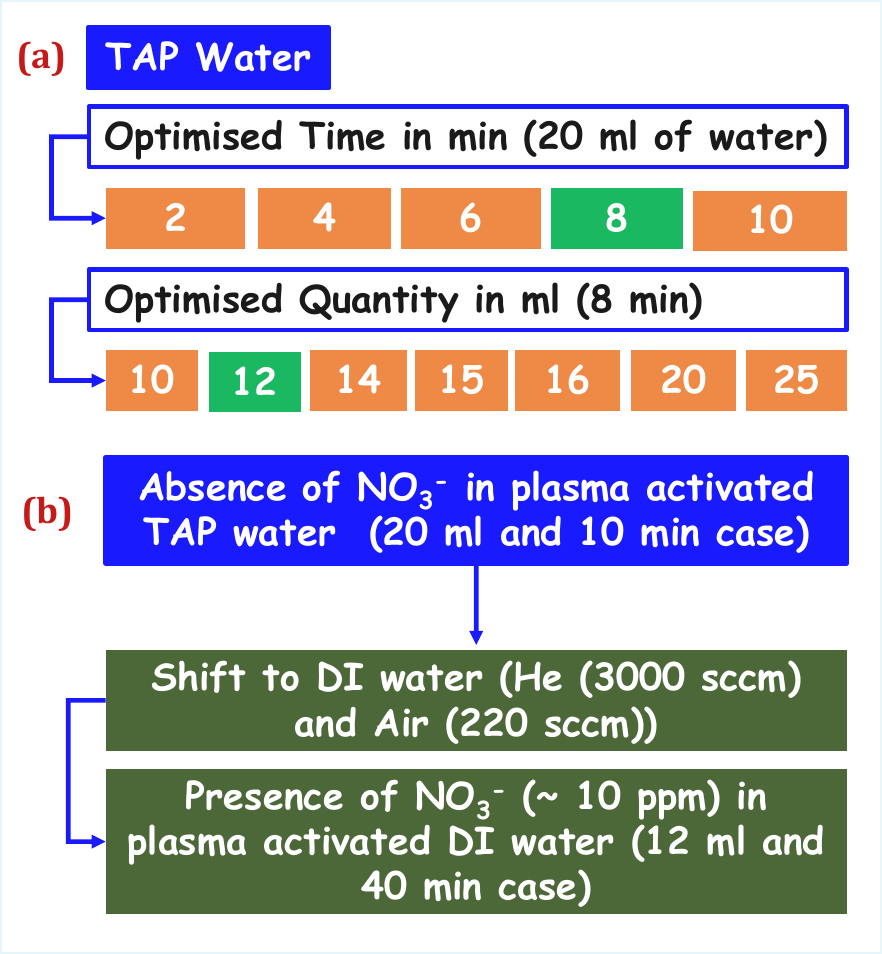}
\caption{Chronology of optimization (a) quantity and treatment time for TAP water  ; green color indicates the optimized quantity and time, and (b) Nitrate presence in plasma-activated  TAP and DI water}
\label{fig:2}
\end{figure}
\normalsize

\begin{figure}
\centering
\includegraphics[scale = 0.55]{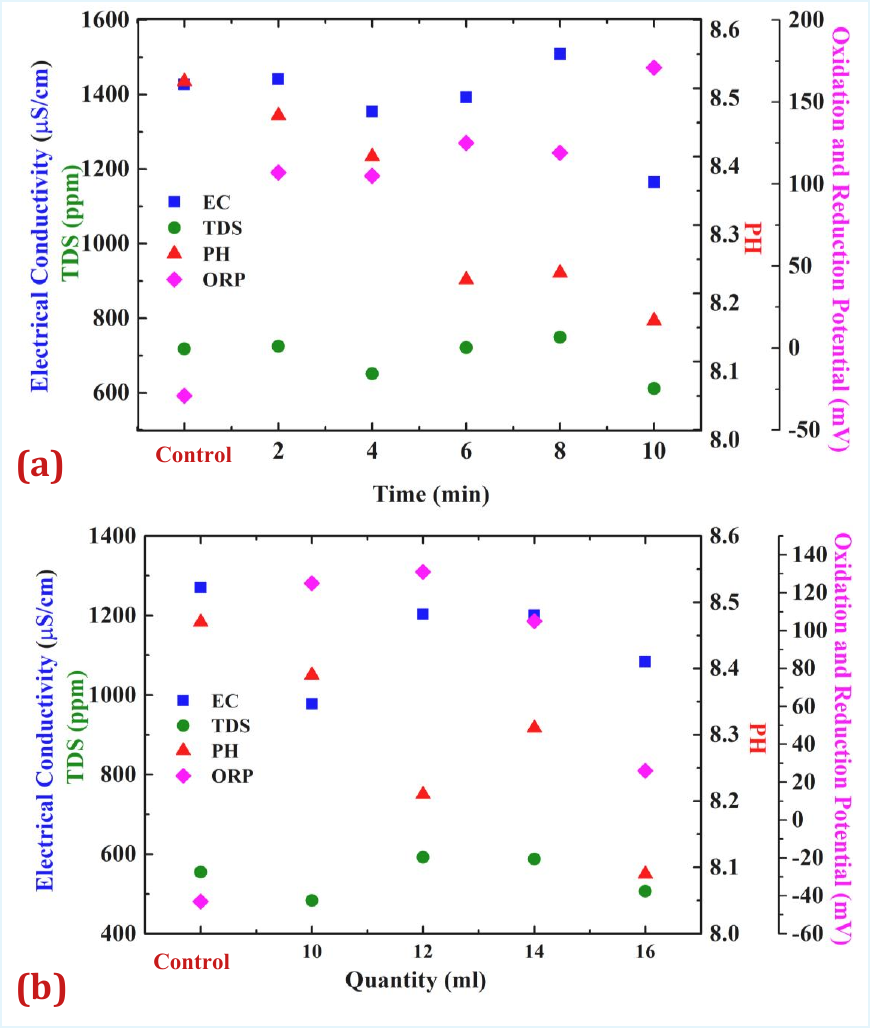}
\caption{Optimization of (a) plasma treatment time for 20 ml of water and (b) water quantity for 8 min treatment time (TAP Water)}
\label{fig:3}
\end{figure}
\normalsize

\section{Optimization process}
The plasma-activated water (PAW) generated was systematically optimized by varying the quantity of water, plasma treatment time, and gas composition (summarized in chronological order in Fig. \ref{fig:2}(a)). Initially, tap water was employed to determine the effect of discharge exposure under controlled laboratory conditions. Optimization of treatment time for 20 ml of tap water was carried out by varying the plasma exposure time from 2 to 10 mins, as shown in Fig. \ref{fig:3}(a). The results show clear variations in pH, electrical conductivity (EC), oxidation-reduction potential (ORP), and total dissolved solids (TDS) with increasing treatment time. Among the tested time durations, the 8 min case was found most suitable as it yielded the maximum value of EC (indicating highest ionic concentration in water), and was therefore selected for subsequent experiments. Thereafter, the quantity of water was optimized while keeping the treatment time fixed at 8 mins (shown in Fig. \ref{fig:3}(b)). The water volumes ranging from 10 ml to 25 ml were treated and analyzed. It was found that the higher water volumes (> 20 ml) led to weaker plasma-liquid interaction and lower chemical activation. Thus, the optimal quantity was found to be 12 ml as it yielded the maximum value of EC and ORP, providing an effective balance between discharge stability and reactive species transfer.

\begin{figure}
\centering
\includegraphics[scale = 0.6]{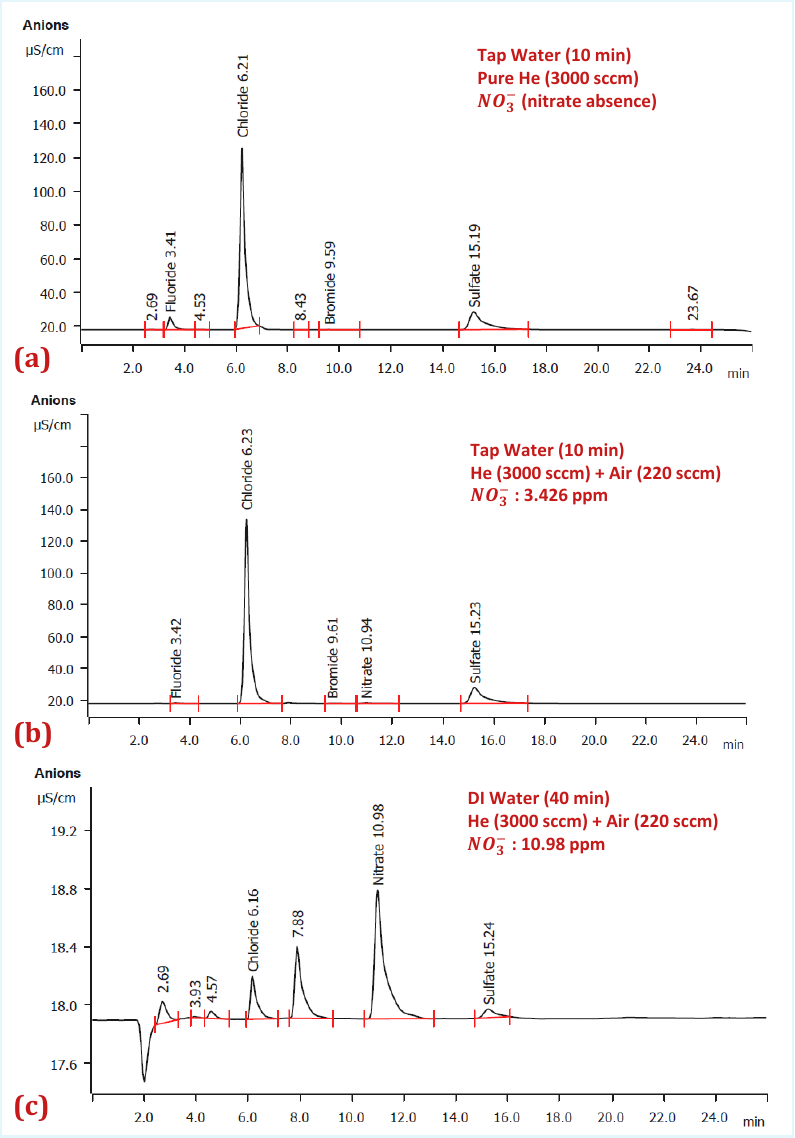}
\caption{Ion chromatography results of nitrate presence for 20 ml of tap water in case of (a) pure helium (3000 sccm), (b) mixture of He (3000 sccm) and Air (220 sccm), and (c) 12 ml of DI water treated with mixture of He (3000 sccm) and Air (220 sccm)
}
\label{fig:4}
\end{figure}

A separate experiment was carried out to test the formation and presence of nitrates upon plasma activation, by employing pure helium (3000 sccm) and readily-available tap water. For this purpose, 20 ml of tap water was treated for 10 mins. The ion chromatography (IC) results revealed a complete absence of nitrate ($NO_3^-$) peaks (as shown in Fig. \ref{fig:4}(a)). This confirmed that pure helium discharge, although visually stable, lacked sufficient nitrogen chemistry to form long-living species such as nitrates or nitrites in the liquid medium. To enhance nitrogen-based reactivity, a small fraction of air (220 sccm) was introduced into the helium flow. A ratio of 14:1 for He to air was found to be optimum considering the maximum amount of air that could be added to the admixture (with the flow rate of He being 3000 sccm), without adversely affecting the discharge stability. A relatively high flow rate of He was found necessary to sustain the plasma jet. Subsequent IC analysis of the helium–air mixture (He 3000 sccm + Air 220 sccm) showed a distinct nitrate peak corresponding to a concentration of $\sim$ 3.4 ppm (shown in Fig. \ref{fig:4}(b)). This shows that even a small admixture of air greatly increases nitrogen excitation and reactive nitrogen species (RNS) formation. Consequently, all subsequent experiments were conducted using this optimized He-Air (14:1) mixture.

\normalsize

\begin{figure}
\centering
\includegraphics[scale = 0.6]{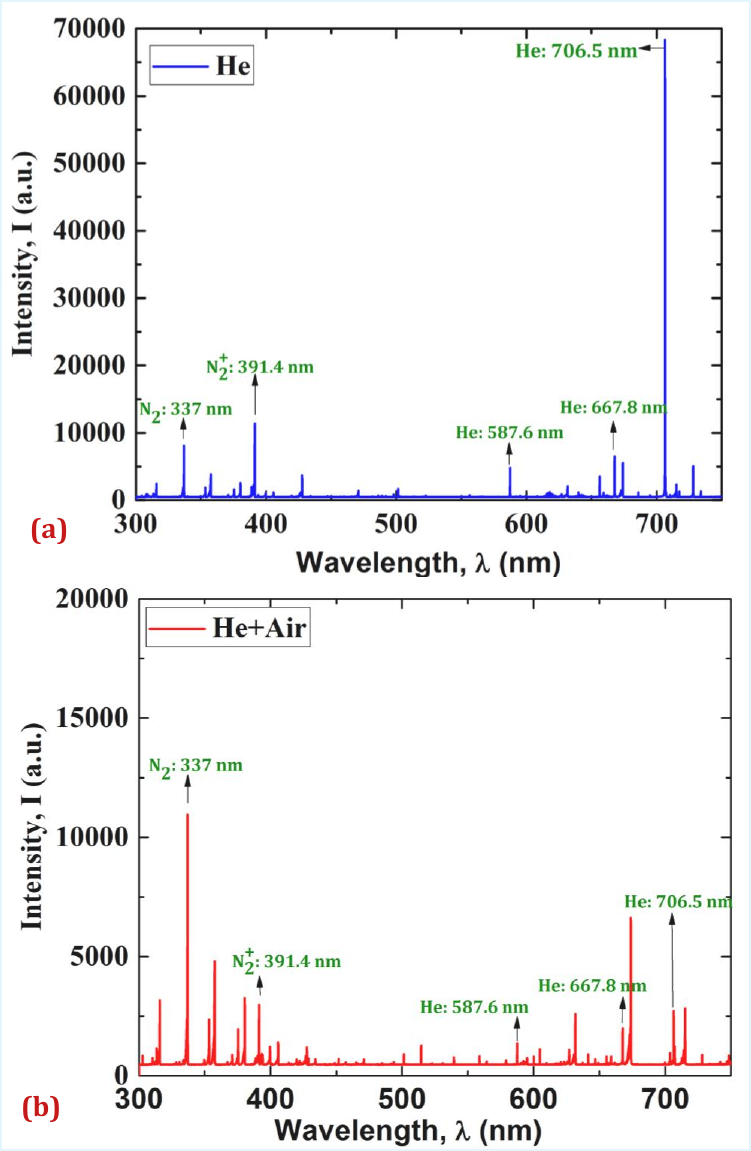}
\caption{Optical Emission spectra of (a) Pure He (3000 sccm) and (b) Mixture of He (3000 sccm) + Air (220 sccm), where some principle lines are mentioned}
\label{fig:5}
\end{figure}
\normalsize

\begin{table}
\caption{\label{tab:1} Plasma Emission Lines}
\begin{ruledtabular}
\begin{tabular}{cccc}
Nature&Species& Wavelength & Transitions\\
 & & (nm) & \\
\hline 
Molecular&$N_2$& 337.1 &$C^2\Pi_u (0)\rightarrow B^3\Pi_g (0)$\\
Ionic&N$_2^+$ & 391.4 &$B^2\Sigma_u^+ (0)\rightarrow X^2\Sigma_g^+(0)$\\
Atomic& He I & 501.6 &$3p^1P^0 \rightarrow 2s^1S$ \\
Atomic& He I & 587.6 &$3d^3D \rightarrow 2p^3P^0$ \\
Atomic& He I & 667.8 &$3d^1D \rightarrow 2p^1P^0$ \\
Atomic& He I & 706.5 &$3s^3S^1 \rightarrow 2p^3P^0$  \\
Atomic& He I & 728.1 &$3s^1S^0 \rightarrow 2p^1P^0$ \\
\end{tabular}
\end{ruledtabular}
\end{table}

\section{Results and Discussion}
\subsection{Optical Emission Spectroscopy (OES) Spectra}

Optical emission spectroscopy (OES) was employed to analyze the emission characteristics of the plasma jet and to qualitatively  identify the excited species responsible for plasma-liquid interactions. The emission spectra of pure helium (3000 sccm) and helium-air mixture (He 3000 sccm + Air 220 sccm) are presented in Fig. \ref{fig:5}(a) and Fig. \ref{fig:5}(b), respectively, and the corresponding principal emission lines and their transitions are summarized in Table \ref{tab:1}. In case of pure helium discharge, the spectrum exhibits strong atomic helium lines at 587.6 nm, 667.8 nm, and 706.5 nm, which correspond to characteristic electronic transitions of helium. A weak nitrogen ($N_2$) second positive system (SPS) band can also be observed in the 337-380 nm region, originating due to trace amounts of nitrogen molecules diffusing from the surrounding ambient air. The overall emission is dominated by helium lines, indicating that the discharge is primarily atomic in nature with limited molecular excitation. After small fraction of air is mixed into the main helium flow, a significant change is observed in the spectral profile. The number of $N_2$ SPS emission lines increases across the 315-400 nm range, although their overall intensity remains comparable to that of the $N_2$ bands seen in the pure helium case. Furthermore, the relative intensity of the helium lines decreases quite a bit, suggesting partial energy transfer from helium metastable to molecular nitrogen and oxygen species. This spectral change indicates the activation of additional excitation channels and the onset of more complex molecular interactions within the discharge.

The emergence of multiple nitrogen bands and the relative suppression of helium lines confirm that the addition of air modifies the discharge composition from a simple atomic plasma to a chemically active molecular plasma. The presence of nitrogen and oxygen species in the plasma plays a pivotal role in generating RONS such as OH, NO, and O, which are known to influence the chemical reactivity of plasma-activated water (PAW). Hence, the He-air (14:1) mixture is identified as a suitable gas composition for producing chemically enriched PAW.

\subsection{Physicochemical Properties and Ion Chromatography (IC) Results of PAW}

\begin{figure}
\centering
\includegraphics[scale = 0.6]{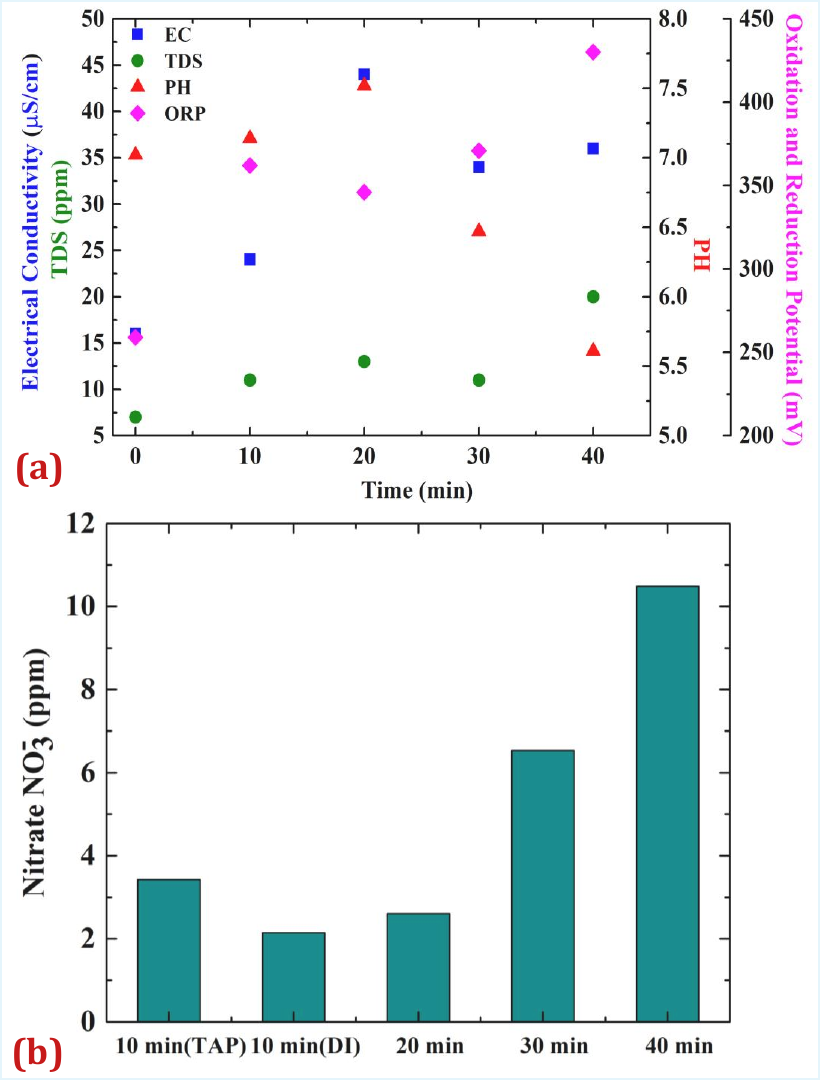}
\caption{Optimization of (a) plasma treatment time for 12 ml of DI water (PAW) and (b) Nitrate concentration (ppm) in DI water with different treatment times along with TAP water (10 min case)}
\label{fig:6}
\end{figure}

The plasma-liquid interaction is further analyzed by evaluating the changes in the physicochemical properties of de-ionized (DI) water after plasma exposure under the optimized discharge conditions. DI water was considered a candidate for the test water sample used, as it inherently had no initial ions and any change in electrical conductivity (EC) could be easily attributed to the generation of new ions in the liquid. Variations in electrical conductivity (EC), oxidation-reduction potential (ORP), total dissolved solids (TDS) and pH as a function of treatment time are shown in Fig. \ref{fig:6}(a). The values of ORP and TDS show a gradual increase with time, while the pH shows a continuous decrease. The EC is seen to gradually increase until about 20 mins of treatment time and it thereafter slightly decreases. This indicates the progressive incorporation of charged and reactive species into the water, such as nitrates ($NO_3^-$) and nitrites ($NO_2^-$). The rise in ORP reflects an enhanced oxidative capacity of the solution, whereas the increase in EC and TDS indicates a higher ionic content, resulting from plasma-induced dissociation and solvation processes. In contrast, the steady decline in pH signifies the acidification of the solution due to the accumulation of acidic nitrogen species ($HNO_2$ and $HNO_3$) \cite{rathore2024innovative}.

Among the tested cases, the 40-minute plasma-treated DI water exhibited the most pronounced changes in physicochemical properties, characterized by the highest ORP and EC along with a moderately acidic pH of around 5.6. This confirms that long-term exposure to plasma increases the chemical reactivity of the water and enhances the formation of long-lived reactive species. The corresponding values of these parameters are consistent with previous reports on atmospheric pressure plasma jets operating in He-air mixtures which also demonstrate that the duration of treatment strongly influences the accumulation of reactive species in PAW \cite{rathore2024innovative,rathore2024effects, wang2025plasma}.  

The chemical composition of the activated water was further examined using ion chromatography (IC), and the results are presented in Fig. \ref{fig:6}(b). The nitrate ($NO_3^-$) concentration, expressed in parts per million (ppm), is measured for different plasma treatment times. A monotonic increase in nitrate concentration with treatment time is observed, which indicates nitrogen fixation processes become more effective under longer plasma exposure. For shorter treatment times (10–20 minutes), the nitrate yield is relatively low. Thereafter, it increases significantly for 30 minutes and beyond. A maximum nitrate concentration of approximately 10 ppm is obtained for the 40-minute treatment of 12 ml DI water (also shown in Fig. \ref{fig:4}(c)), which confirms that this condition corresponds to the most chemically enriched PAW. Based on the above results, the 40-minute plasma-treated de-ionized (DI) water generated using the He-air mixture was identified as the most chemically active condition and was therefore selected for subsequent plant growth studies. Often, a trade-off has to be made regarding the plasma-water activation time, as very long times would result in significant loss of expensive He and prolonged experimental hours in case of multiple runs of the experiment. 

\normalsize

\begin{figure*}
\centering
\includegraphics[scale = 0.55]{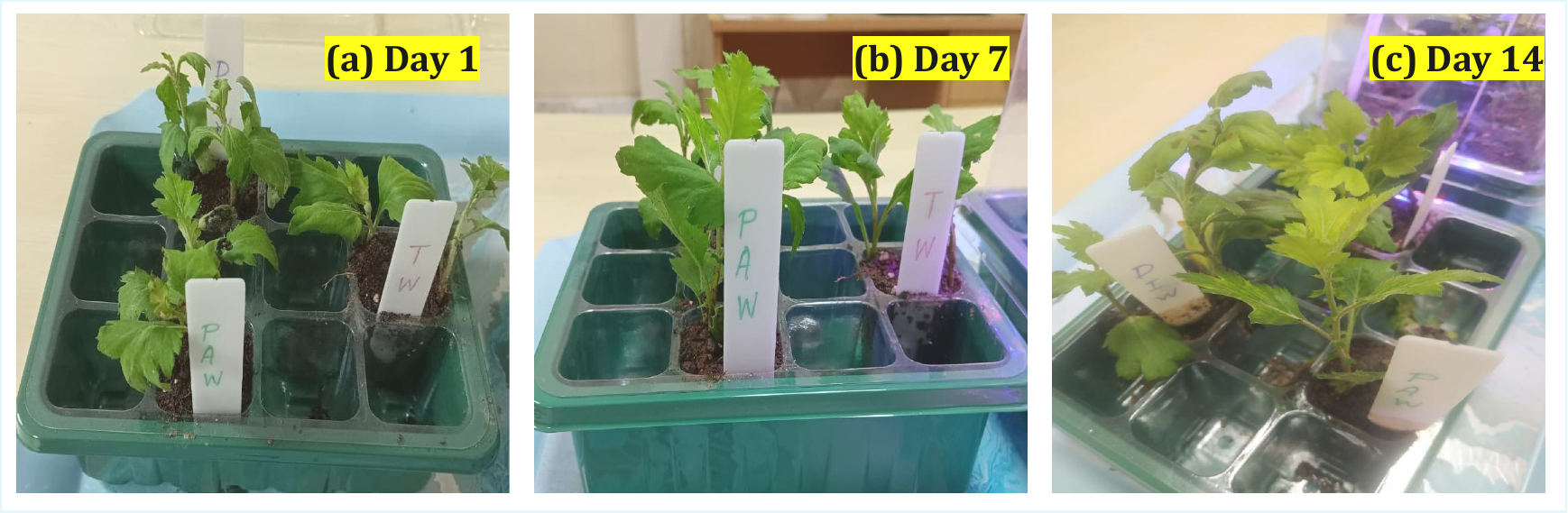}
\caption{Digital pictures of flower sapling growth on (a) Day 1, (b) Day 7, and (c) Day 14}
\label{fig:7}
\end{figure*}
\normalsize

\subsection{Effect of PAW on Flower Sapling Growth}

The optimized plasma-activated water (PAW) generated by treating 12 ml of de-ionized (DI) water for 40 minutes under the He-air mixture discharge is applied to study its effects on the growth of Chrysanthemum saplings. Comparative growth experiments are performed using three cases: (i) PAW (DI water treated for 40 min), (ii) untreated DI water (DIW), and (iii) untreated tap water (TW). As mentioned earlier, each case includes two identical saplings of similar initial height and age. The plants are maintained under identical conditions with controlled illumination and humidity using a Seed Starter Tray, following a 12-hour light and 12-hour dark cycle.

Figures \ref{fig:7} (a), (b) and (c) show the photographs of plant growth on the starting day, after one week, and after two weeks respectively. The average height of the saplings for all cases is summarized in Table \ref{tab:2}, and the corresponding soil fertility measurements are mentioned in Table \ref{tab:3}. As seen in Table \ref{tab:2}, before the experiment, the average heights of the saplings are nearly identical in all cases ($\sim$ 5 cm). After one week, PAW-treated plants displayed faster growth reaching 7.65 cm, compared to 6.25 cm for DIW and 5.45 cm for TW. After two weeks, the PAW-treated saplings achieved an average height of 10.2 cm, significantly higher than the 8.0 cm for DIW and the 7.1 cm for TW. A similar trend was observed in soil fertility, which remained high (2580 $\mu$S/cm) for the PAW-treated pot, but decreased sharply for the DIW (900 $\mu$S/cm) and the TW (795 $\mu$S/cm) cases after 14 days.

The enhanced growth observed for PAW-treated plants can be attributed to the combined effect of RONS and nitrate production during plasma activation. These reactive species are known to improve nutrient absorption, stimulate metabolic activity, and enhance root development. The higher soil fertility further supports the hypothesis that plasma-generated nitrates and related species act as bioavailable nutrients. Overall, the results clearly demonstrate that watering Chrysanthemum saplings with plasma-activated water promotes healthier and faster growth compared to conventional water sources such as tap water, underscoring the viability of PAW as an eco-friendly nutrient source that enhances plant growth and benefits horticultural practices.

\begin{table}
\caption{\label{tab:2}{Average height of plants after 2 weeks for PAW, DIW and TW}}
\begin{ruledtabular}
\begin{tabular}{cccc}
Cases: Avg. Height (cm) & PAW & DIW & TW\\
\hline
Day 1 & 5.35 & 5.3 & 5.1\\
Day 7 & 7.65 & 6.25 & 5.45\\      
Day 14 & 10.2 & 8.0 & 7.1 \\
\end{tabular}
\end{ruledtabular}
\end{table}

\begin{table}
\caption{\label{tab:3}{Soil fertility ($\mu$S/cm) after 2 weeks for PAW, DIW and TW}}
\begin{ruledtabular}
\begin{tabular}{cccc}
Cases: fertility ($\mu$S/cm) & PAW & DIW & TW\\
\hline
Day 1 & 3000 & 3000 & 3000\\
Day 14 & 2580 & 900 & 795 \\
\end{tabular}
\end{ruledtabular}
\end{table}

\section{Conclusion}
A helium-air (14:1) cold atmospheric pressure plasma jet has been utilized to generate plasma-activated water (PAW) and investigate its effect on growth of  Chrysanthemum saplings. Optical emission spectroscopy confirmed the presence of species such as $N_2$, $N_2^+$ and He, indicating a chemically rich environment in the He-air mixture. Ion chromatography analysis verified the formation of nitrate ($NO_3^-$) species in PAW, with the highest concentration of approximately 10 ppm obtained for 12 ml of de-ionized water treated for 40 minutes. The physicochemical characterization of PAW revealed that oxidation-reduction potential (ORP), electrical conductivity (EC), and total dissolved solids (TDS) increased with treatment time, while pH decreased, signifying enhanced chemical reactivity. Comparative plant growth studies demonstrated that PAW-treated Chrysanthemum saplings exhibited a maximum average height of 10.2 cm and soil fertility of 2580 µS/cm, compared to 8.0 cm and 900 µS/cm for deionized water-treated ones (without plasma treatment) and 7.1 cm and 795 µS/cm for tap water-treated ones (without plasma treatment). These results highlight the synergistic effects of reactive species and nitrate enrichment in enhancing plant development, confirming the potential of plasma-activated water as a sustainable nutrient solution for horticultural applications.

\section*{Acknowledgment}
We acknowledge the Department of Science and Technology (DST-SERB), Government of India, for financial support under Grant No. CRG/2022/000112. We also thank the Ion Chromatography Facility of the Department of Chemical Engineering, IIT Kanpur, for providing measurement support and Mr. Sushanta Barman for his assistance during the experiments.
\section*{AUTHOR DECLARATIONS}
\subsection*{Conflict of Interest}
The authors have no conflicts to disclose.
\section*{DATA AVAILABILITY}
The data that supports the findings of this study are available within the article.

\bibliography{bibfile}

\end{document}